\documentclass{PoS}

\title{Conformal or confining -- results from lattice gauge theory for higher-representation gauge theories}

\ShortTitle{Conformal or confining}

\author{\speaker{Benjamin Svetitsky}\\
        Raymond and Beverly Sackler School of Physics and Astronomy, Tel Aviv University, 69978 Tel Aviv, Israel\\
        E-mail: \email{bqs@julian.tau.ac.il}}

\abstract{We have calculated the running coupling in SU(2), SU(3), and SU(4) gauge theories to see whether they have infrared fixed points. An infrared fixed point means no confinement: It means that the long-distance physics is conformal, without a mass scale and indeed without a particle spectrum. All these theories run slowly in perturbation theory, so a nonperturbative determination is interesting. }

\FullConference{Xth Quark Confinement and the Hadron Spectrum\\
		 8--12 October 2012\\
		 TUM Campus Garching, Munich, Germany}

\begin{document}
\def\Tr{{\rm Tr}\,}
\def\eval#1{\left\langle #1 \right\rangle}
\def\co{{\cal O}}
\def\ttl#1{{\em #1}}

\section{Introduction}

For several years my collaborators and I have been studying SU($N$) gauge theories, $N=2,3,4$,
with fermions in the two-index symmetric representation of color
\cite{Shamir:2008pb}--\cite{DeGrand:2012qa}.
In each case, we have examined the theory with 2 flavors.
As a function of the number of flavors, these theories lie near the borderline between confinement
and screening~\cite{Dietrich:2006cm}; screened gauge theories are believed to be governed by an infrared fixed
point, and hence are called {\em conformal}.
The main question we have been asking is whether each theory is above or below the borderline,
that is, whether it is inside the {\em conformal window\/} in $N_f$.
With results for these three theories in hand, it is a good time to sum things up.

QCD, familiar and much-studied, has a rapidly running coupling that becomes
strong in the infrared, where the consequences are quark confinement and spontaneous breaking
of chiral symmetry.
The monotonic running of the coupling is expressed by the strictly negative $\beta$ function of QCD\@.
As we increase the number of massless flavors in QCD, we expect (based on the form of the two-loop
$\beta$ function~\cite{Caswell:1974gg,Banks:1981nn}) that the $\beta$ function will develop a shape that includes a zero-crossing
at some coupling $g^*$; this is an infrared fixed point, where the theory is trapped as we go
to large distance scales.
Of special interest is the regime in $N_f$ where the theory is just short of forming an IRFP;
this is a theory with a $\beta$ function that lies near the axis, a {\em walking\/} theory.

Our approach to the theories at hand is via their $\beta$ functions.
We use the Schr\"odinger functional method \cite{Luscher:1992an,Luscher:1993gh} to calculate the running coupling and to see how
it changes with scale; the scale chosen is the volume of the lattice, which affects the coupling
through a background field.
No lattice calculation, however, can proceed without understanding the phase diagram of the
theory, and indeed the phase diagram offers its own hints about the physics of the
continuum theory.
Finally, once the Schr\"odinger functional is implemented, it is straightforward to go beyond
the $\beta$ function to calculate the anomalous dimensions of various operators.
The anomalous dimension $\gamma_m$ of the fermion mass parameter \cite{Bursa:2009we} is in fact easier to calculate
than the $\beta$ function, and it is not without phenomenological interest.
I will describe the method and results in this talk.
But first I'd like to make some general comments about the difficulty of life near the sill of the
conformal window. 

The cases of walking and of an IRFP are both hard cases for study, and they are hard to
tell apart.
They are characterized by slow running (at best), so that strong coupling in the IR---where
we look for the fixed point---is also strong coupling in the UV---that is, at the scale
of the lattice spacing $a$.
A QCD-style weak-coupling continuum limit would require an enormous ratio between the size $L$
of the lattice and the lattice spacing $a$.
On the other hand, any lattice of ordinary, practical dimensions allows you to look only at a
narrow range of scales.
In fact, even in a clearly conformal theory the largest scale $L$ may still leave you far from the
IRFP, so that you miss the scale invariance entirely.

Another way of saying this \cite{DeGrand:2009hu} is to note that if there is an IRFP then the gauge coupling is
an irrelevant coupling under the RG, while the quark mass $m_q$ is a relevant parameter, as is $1/L$.
This means that the massless limit would bring into play very bad finite-size effects.

All this supplies motivation to use the Schr\"odinger functional, which turns finite volume from
a hindrance into a method.

\section{The $\beta$ function in the massless theory}

The Schr\"odinger functional method defines the running coupling as follows.
Take a hypercubical Euclidean box of volume $L^4$ and a massless gauge theory.
We fix the gauge field $A^i({\bf x},t)$ on the two time boundaries of the box, $t=0,L$.
This sets up a background electric field in the box; this field is the unique minimum of
the classical action $S_{\textrm{cl}}=\int d^4x \,F_{\mu\nu}^2$.
We scale the field with $L$ so that $L$ is the {\em only\/} scale in the problem.
Now we calculate (in principle) the quantum partition function $Z$ and, from it, the
effective action $\Gamma=-\log Z$ of the background field and its quantum fluctuations.
In  perturbation theory it has to be proportional to the classical action, and the constant of
proportionality is just the running coupling at the scale $L$,
\begin{equation}
\Gamma=\frac1{g^2(L)}S_{\textrm{cl}}.
\end{equation}
The SF method takes this to be the {\em nonperturbative\/} definition of the running coupling:
It is the ratio of $\Gamma$ to the classical action of the background field.

Let me skip most of the details of how we carry out such a calculation on the lattice.
The most important fact is that, since $\Gamma$ cannot be calculated directly, we work with
its derivative $d\Gamma/d\eta$ with respect to some parameter $\eta$ in the boundary gauge field.
This can be calculated as a lattice correlation function, and we then compare its value
to the same derivative of $S_{\textrm{cl}}$ to obtain $g^2(L)$.

We work with Wilson--clover fermions, using fat links in the fermion action to suppress
lattice artifacts.
For each value of the (bare) lattice coupling $g_0$, we find the critical hopping parameter
$\kappa_c(g_0)$ (related to the bare mass) by demanding that the current mass $m_q$ of the
fermion be zero.
The current mass is defined by an axial Ward identity that relates correlation functions
of the axial current to those of the pseudoscalar density,
\begin{equation}
\partial_\mu\langle A^{a\mu}(x)\co^a\rangle=2m_q\langle P^a(x)\co^a\rangle,
\label{AWI}
\end{equation}
where $\co^a$ is a convenient source on a wall of the lattice.
Then, given $g_0$ and $\kappa$, we calculate the running coupling for a range of
lattice sizes $L=6a,\ldots,16a$.

\begin{figure}[bt]
\begin{center}
\includegraphics*[width=.40\columnwidth]{1g2_vs_l_bqs.eps}\ \ \ \ 
\includegraphics*[width=.45\columnwidth]{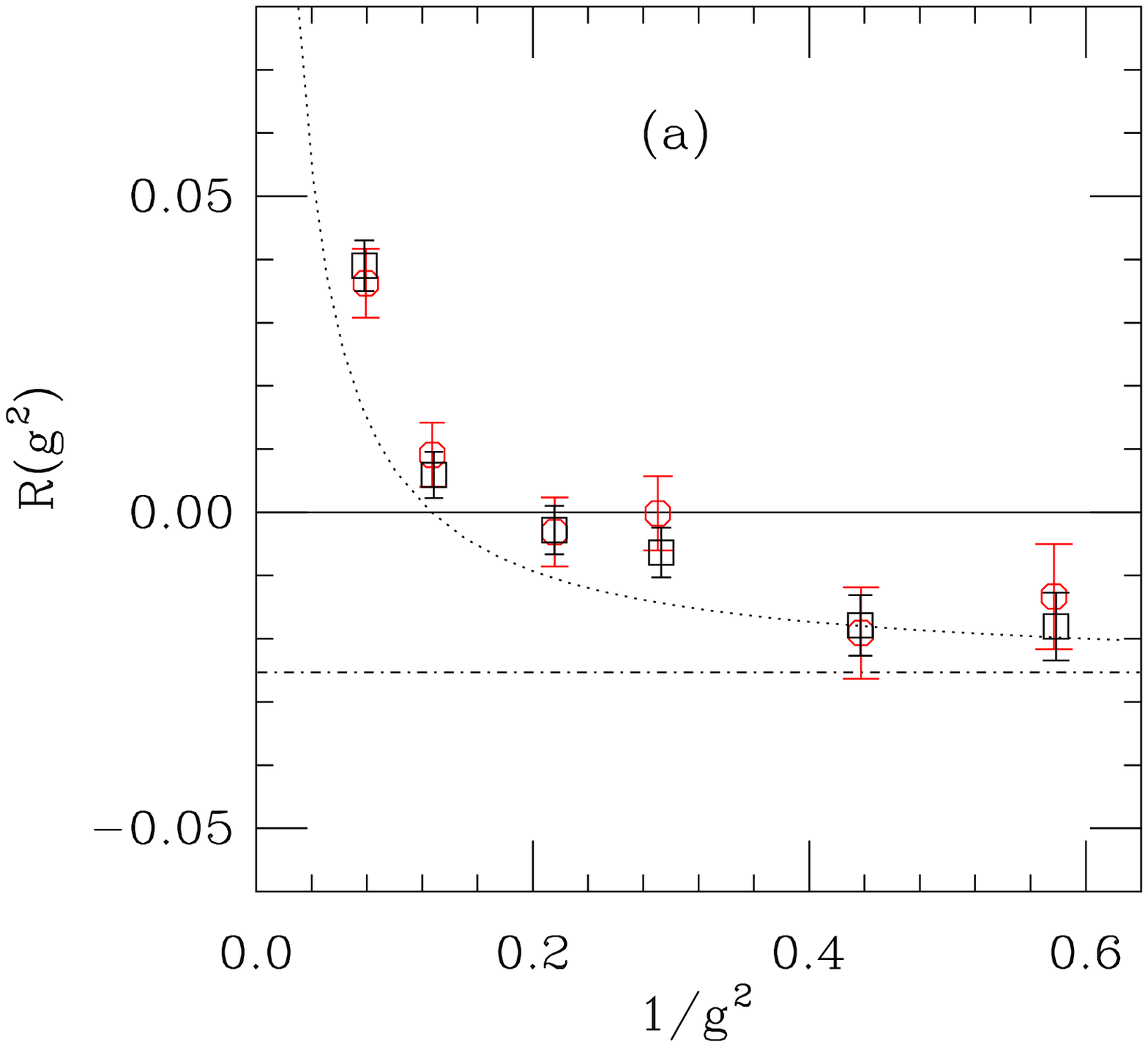}
\caption{SU(2)/triplet theory \cite{DeGrand:2011qd}. Left: SF coupling vs.~size  $L$ of the lattice, for bare couplings $g_0$ ranging from weak (top) to strong (bottom).
The fitted slopes give the $\tilde\beta$ function.
The dotted line has the slope of $\tilde\beta$ calculated in the weak coupling limit (one loop).
Right:  The $\tilde\beta$ function extracted from these slopes.
The dash-dotted line is the one-loop value and the dotted curve is the two-loop result.
\label{SU2coupling}}
\end{center}
\end{figure}

Let's focus now on the SU(2) theory~\cite{DeGrand:2011qd}. (The fermions are color triplets.)
If we plot the running coupling against $\log L$ (Fig.~\ref{SU2coupling})
we find that the data for each value of $g_0$ line up.
Starting with $L=6a$, let's define a variable for the inverse coupling at scale
factor $s$ according to
$u(s)\equiv1/g^2(sL)$.
The beta function for $u$ is
\begin{equation}
\tilde\beta(u)\equiv du/d\log s=2\beta(g^2)/g^4.
\end{equation}
The slopes in Fig.~\ref{SU2coupling} would give $\tilde\beta(u)$ directly, except that the
running coupling $g^2(L)$ changes as $L$ is increased for any $g_0$.
Here, however, we can take advantage of the fact that the theory is either conformal or
close to it: The running of the coupling is so slow that the coupling changes little along
the lines drawn in the figure, and hence $\tilde\beta(u)$---the slope of each line---also changes little.
Extracting the slopes and plotting them against the coupling $g(L=8a)$ gives the approximate
$\tilde\beta(u)$ plotted in the right-hand figure.
Evidently the $\beta$ function crosses zero, an infrared fixed point.
Interestingly, the lattice result tracks the two-loop $\beta$ function fairly well.
Comparison of the black squares (the full fits) to the red circles (obtained by dropping the $L=6a$ lattices)
gives some idea of the sensitivity of the result to the lattice spacing.

\section{Phase Diagram}

\begin{figure}[bt]
\begin{center}
\includegraphics*[width=.5\columnwidth]{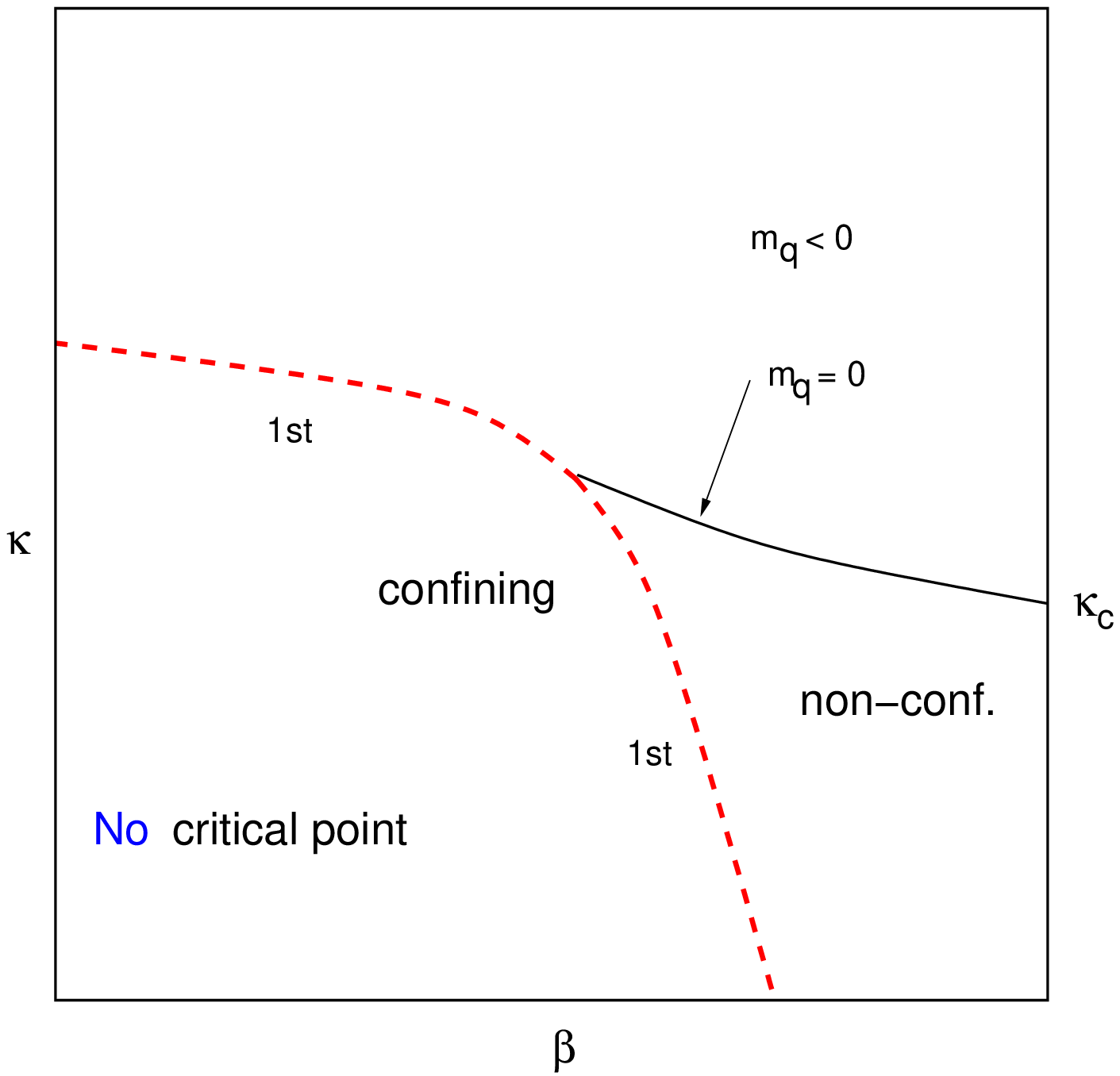}
\caption{Sketched phase diagram for the SU(3)/sextet theory.
\label{phasediagram}}
\end{center}
\end{figure}

Let's pause for a look at a phase diagram.
Fig.~\ref{phasediagram} is a sketch of the phase diagram we've explored \cite{DeGrand:2010na} for the SU(3) theory
(with sextet fermions) on a finite lattice.
The lower left-hand corner of the diagram is a strong-coupling, confining phase, walled off by
(1) the first-order finite-temperature transition to the right, which connects to (2) a first-order
phase boundary reached by increasing $\kappa$ in the small $\beta$($\equiv6/g_0^2$) regime.
There are two important consequences of this diagram for our work.
\begin{enumerate}
\item How is one to contemplate a continuum limit for this theory?
In QCD, the horizontal part of the phase boundary is a second-order transition to the Aoki
phase.
One reaches a theory with massless pions on this curve; then, as $L$ is increased, the corner of
the confining phase slides to the right, towards a continuum limit with a massless (or at least a light) pion.
In the sextet theory, on the other hand, the confining phase is strictly massive and so is the phase boundary; it is hard to imagine taking such
a limit, unless there is a dramatic change in the phase boundary when $L$ is sufficiently large.
The alternative is to take a continuum limit in the non-confining phase on the right, on the
$\kappa_c(\beta)$ curve where $m_q=0$.
Is this a sign of a conformal infrared theory?
(This conundrum was posed long ago \cite{Iwasaki:1991mr,Iwasaki:2003de} for QCD with $N_f\gtrsim8$.  See also \cite{Nagai:2009ip}.)
\item For the SF method we require $m_q=0$.
This can only be achieved in the right side of the diagram.
As we go towards stronger coupling in search of an IRFP, we will be forced to halt where
the $\kappa_c$ curve hits the phase boundary, since there is no locus where $m_q=0$ to the left
of that point.
($m_q$ changes sign discontinuously on the horizontal phase boundary.)
\end{enumerate}

The second point is the reason that we cannot continue measuring the $\beta$ function plotted
in Fig.~\ref{SU2coupling} at yet stronger couplings; we are lucky to be able to reach the IRFP in the SU(2)/triplet theory.
In the SU(3)/sextet and SU(4)/decuplet theories the consequences are yet more serious, since the phase boundary is reached at relatively weak couplings, well short of the two-loop fixed point in each case.
Our solution to this problem is to generalize the lattice action to include a pure-gauge plaquette term built out of the fat links used in the fermion term,
\begin{equation}
S_g=\frac{\beta}{2N_c}\sum\Tr U_p+\frac{\beta_f}{2d_f}\sum\Tr V_p\ .
\end{equation}
Judicious choice of $\beta_f$ moves the phase boundary and thus expands the accessible range of the running coupling.

\begin{figure}[bt]
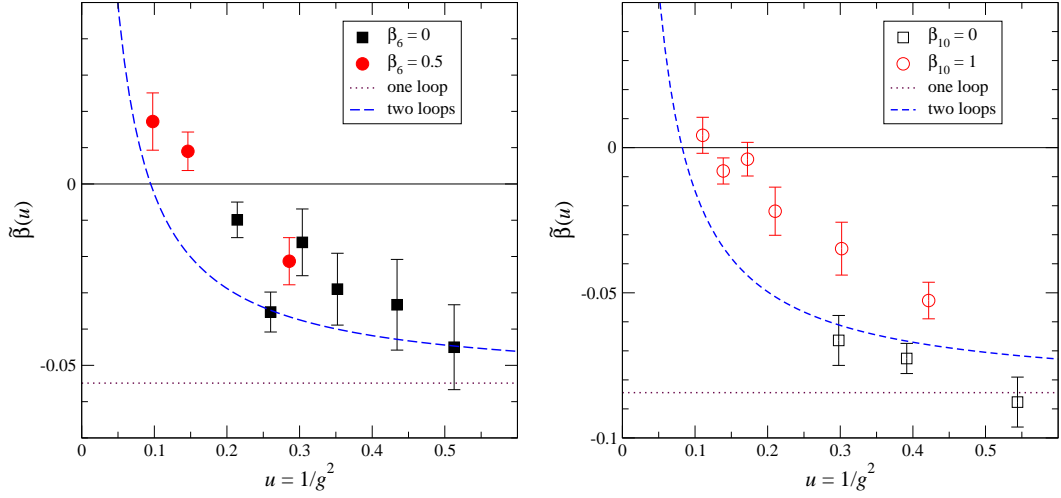

\begin{center}
\includegraphics*[width=.45\columnwidth]{SU3betacombined4.eps}\ \ \ \ 
\includegraphics*[width=.45\columnwidth]{SU4beta1.eps}
\caption{$\tilde\beta$ functions for the SU(3)/sextet (left) and SU(4)/decuplet (right) theories
\cite{DeGrand:2012yq,DeGrand:2012qa}.
Black squares vs.~ red circles denote the two different actions used for each theory.
\label{SU34beta}}
\end{center}
\end{figure}

I show the $\tilde\beta$ functions for the SU(3) \cite{DeGrand:2012yq} and SU(4) \cite{DeGrand:2012qa} theories in Fig.~\ref{SU34beta}.
It looks like the SU(3) theory possesses an IRFP, but the error bars on the leftmost points don't allow this conclusion to be stated with confidence.
It is quite possible that the $\beta$ function turns away from the axis without crossing it, as envisioned in a ``walking'' scenario.
This is true all the more for the SU(4) theory.
It is important to note that we cannot yet provide a systematic continuum limit for the $\beta$ 
function in any of these theories because large fluctuations prevent a smooth extrapolation.

\section{Mass anomalous dimension}

From the axial Ward identity (\ref{AWI}), one derives that the anomalous dimension $\gamma_m$ of the mass parameter can be extracted from the normalization $Z_P$ of the pseudoscalar density $P^a(x)$.
We obtain $Z_P$ from a correlation function,
\begin{equation}
\left.\eval{P^b(t)\; \co^b(t'=0)}\right|_{t=L/2}=Z_P\,Z_\co\, e^{-m_\pi L/2},
\end{equation}
which propagates the lightest particle in the theory from a source $\co$ on the wall to the center of the lattice; we remove the normalization $Z_\co$ of $\co$ as well as $m_\pi$ by calculating a wall-to-wall correlator,
\begin{equation}
\eval{\co^b(t=L)\; \co^b(t'=0)}=Z_\co^2\, e^{-m_\pi L}\ .
\end{equation}
Upon extracting $Z_P$, we use its $L$-dependence to find $\gamma_m$ via
\begin{equation}
\frac{Z_P(L)}{Z_P(L_0)}=\left(\frac{L}{L_0}\right)^{-\gamma_m},
\end{equation}
which again incorporates the observation that the running is slow and hence $\gamma_m(g^2)$ is
approximately constant as $L$ changes at fixed $g_0$.

\begin{figure}[bt]
\begin{center}
\includegraphics*[width=.45\columnwidth]{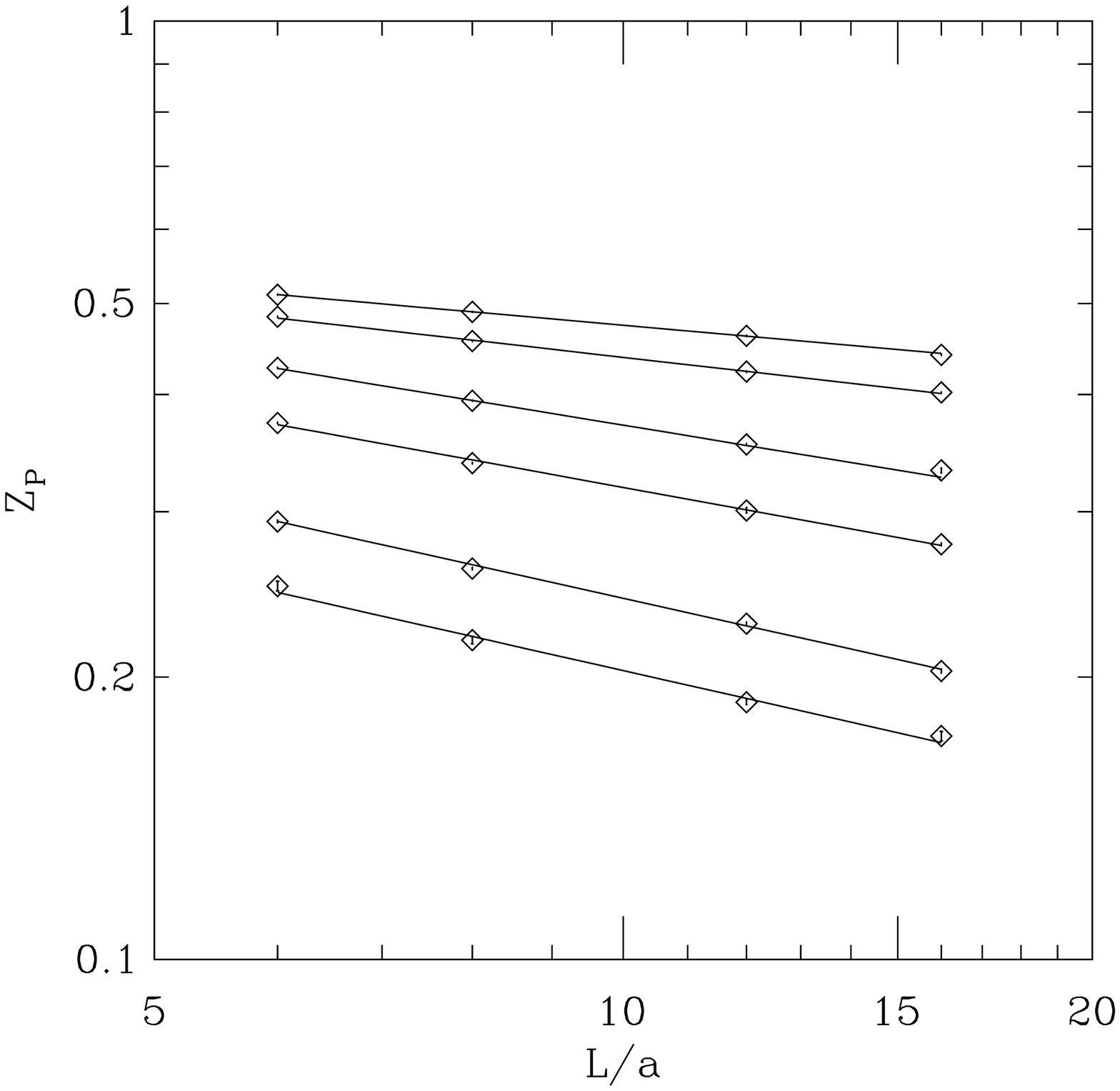}\ \ \ \ 
\includegraphics*[width=.45\columnwidth]{gammalargeN_bqs.eps}
\caption{Left: SU(2)/triplet theory \cite{DeGrand:2011qd}. Pseudoscalar renormalization constant $Z_P$ plotted against the size $L$ of the lattice, for bare couplings $g_0$ ranging from weak (top) to strong (bottom).
The fitted log--log slopes give $-\gamma_m$.
Right: $\gamma_m(g^2)$ for the SU(2,3,4) theories, plotted together~\cite{DeGrand:2012qa}.
\label{SU2Z}}
\end{center}
\end{figure}

\begin{figure}[bt]
\begin{center}
\includegraphics*[width=.5\columnwidth]{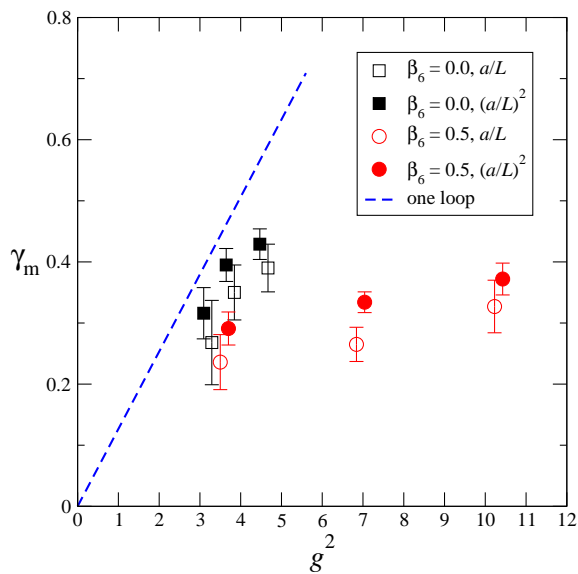}
\caption{SU(3)/sextet theory \cite{DeGrand:2012yq}: Continuum extrapolations of $\gamma_m(g^2)$.
the black squares  and red circles originate from two different lattice actions, while empty (filled) symbols represent linear (quadratic) extrapolations.
\label{extrap}}
\end{center}
\end{figure}

The extraction of $\gamma_m(g^2)$ from a log--log fit is illustrated in Fig.~\ref{SU2Z}.
I've plotted there the results for all three theories, using the large-$N$ variable $g^2N$ as
a common abscissa \cite{DeGrand:2012qa}.
This plot is remarkable for three reasons:
\begin{enumerate}
\item The curves for all three theories depart from the one-loop line and level off.
\item They level off at about the same value.
\item This value is much less than one.
\end{enumerate}
The last point is significant because various approximations and models \cite{Yamawaki:1985zg,Cohen:1988sq,Kaplan:2009kr} predict that theories near the sill of the conformal window, and particularly walking theories, should give $\gamma_m\simeq1$.
Once more, however, I have plotted results that do not take account of finite lattice error.
Since many people are upset by the finding that $\gamma_m$ never approaches 1, we have invested extra effort and produced a continuum extrapolation for the SU(3)/sextet theory \cite{DeGrand:2012yq}.
As can be seen in Fig.~\ref{extrap}, the error bars have grown and there is added uncertainty in how to extrapolate, but on the whole the extrapolations drive $\gamma_m$ further down.

\section*{Acknowledgments}

My collaborators in all this work have been Tom DeGrand and Yigal Shamir, but some interpretation presented here is my own.
This work was supported by the US Department of Energy and by the Israel Science Foundation.
Computer time was granted by the US National Science Foundation through the TeraGrid program and by the US Department of Energy through the USQCD collaboration.
Our software grew out of the publicly available code of the MILC collaboration.


\begin{thebibliography}{99}

\bibitem{Shamir:2008pb}
  Y.~Shamir, B.~Svetitsky and T.~DeGrand,
  \ttl{Zero of the discrete beta function in SU(3) lattice gauge theory with color sextet fermions,}
  Phys.\ Rev.\  D {\bf 78}, 031502 (2008)
  [arXiv:0803.1707 [hep-lat]].

\bibitem{DeGrand:2008kx}
  T.~DeGrand, Y.~Shamir and B.~Svetitsky,
  \ttl{Phase structure of SU(3) gauge theory with two flavors of symmetric-representation fermions,}
  Phys.\ Rev.\  D {\bf 79}, 034501 (2009)
  [arXiv:0812.1427 [hep-lat]].

\bibitem{DeGrand:2010na}
  T.~DeGrand, Y.~Shamir, B.~Svetitsky,
  \ttl{Running coupling and mass anomalous dimension of SU(3) gauge theory with two flavors of symmetric-representation fermions,}
  Phys.\ Rev.\  {\bf D82}, 054503 (2010).
  [arXiv:1006.0707 [hep-lat]].
  
\bibitem{DeGrand:2011qd}
  T.~DeGrand, Y.~Shamir, B.~Svetitsky,
  \ttl{Infrared fixed point in SU(2) gauge theory with adjoint fermions,}
  Phys.\ Rev.\  {\bf D83}, 074507 (2011).
  [arXiv:1102.2843 [hep-lat]].
  
\bibitem{DeGrand:2012yq} 
  T.~DeGrand, Y.~Shamir and B.~Svetitsky,
  \ttl{Mass anomalous dimension in sextet QCD,}
  arXiv:1201.0935 [hep-lat].
  
\bibitem{DeGrand:2012qa} 
  T.~DeGrand, Y.~Shamir and B.~Svetitsky,
  \ttl{SU(4) lattice gauge theory with decuplet fermions: Schr\"odinger functional analysis,}
  Phys.\ Rev.\ D {\bf 85}, 074506 (2012)
  [arXiv:1202.2675 [hep-lat]].

\bibitem{Dietrich:2006cm} 
  D.~D.~Dietrich and F.~Sannino,
  \ttl{Conformal window of SU(N) gauge theories with fermions in higher dimensional representations,}
  Phys.\ Rev.\ D {\bf 75}, 085018 (2007)
  [hep-ph/0611341].
  
\bibitem{Caswell:1974gg}
  W.~E.~Caswell,
  \ttl{Asymptotic behavior of nonabelian gauge theories to two loop order,}
  Phys.\ Rev.\ Lett.\  {\bf 33}, 244 (1974).

\bibitem{Banks:1981nn}
  T.~Banks and A.~Zaks,
  \ttl{On the phase structure of vector-like gauge theories with massless fermions,}
  Nucl.\ Phys.\  B {\bf 196}, 189 (1982).

\bibitem{Luscher:1992an}
  M.~L\"uscher, R.~Narayanan, P.~Weisz and U.~Wolff,
  \ttl{The Schrodinger functional: A Renormalizable probe for nonAbelian gauge
  theories,}
  Nucl.\ Phys.\  B {\bf 384}, 168 (1992)
  [arXiv:hep-lat/9207009].

\bibitem{Luscher:1993gh}
  M.~L\"uscher, R.~Sommer, P.~Weisz and U.~Wolff,
  \ttl{A precise determination of the running coupling in the SU(3) Yang-Mills theory,}
  Nucl.\ Phys.\  B {\bf 413}, 481 (1994)
  [arXiv:hep-lat/9309005].

\bibitem{Bursa:2009we}
  F.~Bursa, L.~Del Debbio, L.~Keegan, C.~Pica and T.~Pickup,
  \ttl{Mass anomalous dimension in SU(2) with two adjoint fermions,}
  Phys.\ Rev.\  D {\bf 81}, 014505 (2010)
  [arXiv:0910.4535 [hep-ph]].

\bibitem{DeGrand:2009hu}
  T.~DeGrand,
  \ttl{Finite-size scaling tests for SU(3) lattice gauge theory with color sextet fermions,}
  Phys.\ Rev.\  D {\bf 80}, 114507 (2009)
  [arXiv:0910.3072 [hep-lat]].

\bibitem{Iwasaki:1991mr}
  Y.~Iwasaki, K.~Kanaya, S.~Sakai and T.~Yoshi\'e,
  \ttl{Quark confinement and number of flavors in strong coupling lattice QCD,}
  Phys.\ Rev.\ Lett.\  {\bf 69}, 21 (1992).

\bibitem{Iwasaki:2003de}
  Y.~Iwasaki, K.~Kanaya, S.~Kaya, S.~Sakai and T.~Yoshi\'e,
  \ttl{Phase structure of lattice QCD for general number of flavors,}
  Phys.\ Rev.\  D {\bf 69}, 014507 (2004)
  [arXiv:hep-lat/0309159].

\bibitem{Nagai:2009ip}
  K.~Nagai, G.~Carrillo-Ruiz, G.~Koleva and R.~Lewis,
  \ttl{Exploration of SU($N_c$) gauge theory with many Wilson fermions at strong coupling,}
  Phys.\ Rev.\  D {\bf 80}, 074508 (2009)
  [arXiv:0908.0166 [hep-lat]].

\bibitem{Yamawaki:1985zg}
  K.~Yamawaki, M.~Bando and K.~i.~Matumoto,
  \ttl{Scale Invariant Technicolor Model And A Technidilaton,}
  Phys.\ Rev.\ Lett.\  {\bf 56}, 1335 (1986).

\bibitem{Cohen:1988sq} 
  A.~G.~Cohen and H.~Georgi,
  \ttl{Walking Beyond The Rainbow,}
  Nucl.\ Phys.\ B {\bf 314}, 7 (1989).
  
\bibitem{Kaplan:2009kr} 
  D.~B.~Kaplan, J.~-W.~Lee, D.~T.~Son and M.~A.~Stephanov,
  \ttl{Conformality Lost,}
  Phys.\ Rev.\ D {\bf 80}, 125005 (2009)
  [arXiv:0905.4752 [hep-th]].

  
\end{thebibliography}
\end{document}